\documentclass{aa}
\usepackage{epsfig}
\begin{document}

   \thesaurus{06     
              (08.14.1; 08:16:6; 09:19:1)}  
%
   \title{Search for Discrete Refractive Scattering Events }


   \author{R. Ramachandran$^1$, A. A. Deshpande$^{2,3}$, B. W. Stappers$^4$} 
   \offprints{ramach@astro.uva.nl}
   \institute{$^1$Netherlands Foundation for Research in Astronomy, Postbus 2,
              7990 AA Dwingeloo, The Netherlands \\
              $^2$Raman Research Institute, C. V. Raman Avenue, 
              Bangalore - 560 080, India\\
              $^3$Austarlia Telescope National Facility, CSIRO, Narrabri, 
              NSW, Australia \\
              $^4$Sterrenkundig Instituut, Universiteit van Amsterdam, NL-1098 SJ
              Amsterdam, The Netherlands. \\
              email:ramach@astro.uva.nl; desh@rri.ernet.in; bws@astro.uva.nl}
   \date{}
   \authorrunning{R. Ramachandran, A. A. Deshpande \& B. W. Stappers}
   \titlerunning{Search for Single Scattering Events}
   \maketitle

\begin{abstract} 
We have searched for discrete refractive scattering events (including effects
due to possible non-multiple diffractive scattering) at meter wavelengths in the
direction of two close by pulsars B0950+08 and B1929+10, where we looked for
spectral signatures associated with the multiple imaging of pulsars due to
scattering in the interstellar medium.  We do not find any signatures of such
events in the direction of either source over a spectral periodicity range of 50
KHz to 5 MHz.  Our analysis puts strong upper limits on 
the column density contrast associated with a range of spatial scales of
the interstellar electron density irregularities along these lines of sight.
\end{abstract} 

\section{Introduction}
Radio signals, during their passage through the interstellar medium, are
scattered due to the irregularities in the density of free electrons (Scheuer
1968). Most often, signals from distant sources undergo strong and multiple
scattering. This leads to many observable effects such as the apparent angular
broadening of the source, temporal broadening of the pulse profile, diffractive
and refractive scintillations. The nature of interstellar scattering at this
limit and the effect on relevant observables have been studied in detail by many
early workers (Lee \& Jokipii 1975; 1976; Blandford \& Narayan 1985; Blandford,
Narayan \& Romani 1984; Cordes 1986). Refractive alterations (by irregularities
larger than the relevant Fresnel scale) of diffraction patterns sometimes lead
to multiple images with angular separations larger than the diffraction broadening
of the source (Cordes \& Wolsczan 1986; Cordes et al. 1986; Goodman \& 
Narayan 1989a,b). 

If the refractive alterations become significant, then multiple imaging
manifests itself in the form of periodic spectral \& temporal modulation of
intensity with periods smaller than the decorrelation bandwidth \& decorrelation
time, respectively. The probablity of multiple imaging occuring is low for a
Kolmogorov spectrum of density distribution, and increases for steeper spectra
as they imply dominant refractice effects (see Cordes et al. 1986 for details).
However, for nearby sources, refraction angles may become comparable with
the diffraction broadening even for relatively less steep density spectra.  It is
also likely that the diffractive scattering of the signals from some of the
close-by sources is less `multiple' in nature (with a relatively small number of
speckles), mimicing in some sense the multiple imaging situation, where, with an
unscattered version of the signal, only a few delayed versions interfere.  Also,
this may improve the detectability of a distinct signature of {\it single} or {\it
discrete refractive scattering events}, where we receive, only a few discrete
(diffractive) bundles of rays that correspond to different refracted (and
correspondingly) delayed versions of the signal.  In such a case, it appears
possible to probe the properties (such as the size and density contrast) of the
discrete density-irregularities responsible for the refractive/diffractive
scattering, if the associated time delays can be measured.

\section{Observations and Analysis}
\label{sec-analysis}
We observed two near-by pulsars, B0950+08 and B1929+10 with the aim of detecting
discrete refractive scattering events. 
These observations were carried out with the
Westerbork Synthesis Radio Telescope (WSRT) with its pulsar backend, {\tt
PuMa}. The WSRT consists of 14 dishes, each of 25-m diameter. For this
observation, the delays between the dishes were compensated, and the signals
were added in phase to construct an equivalent single dish of about 94 m
diameter, with an antenna gain of about 1.2 K/Jy. Observations were conducted on
1999 Feb 22, and 1999 Apr 28 at a centre frequency of 382 MHz, with a bandwidth
of 10 MHz. The signal voltages were Nyquist-sampled and recorded at 20 MHz 
in both X and Y polarisation channels. The data are quantized to represent each
sample with 2 bits.

During the off-line analysis, we first performed ``coherent de-dispersion''
(Hankins 1971), to remove the effect of interstellar dispersion on the signals
corresponding to the X and the Y polarisation channels. The assumed dispersion
measure values were, $2.9704\pm 0.0001$ pc cm$^{-3}$ for PSR B0950+08, and
$3.1760\pm 0.003$ pc cm$^{-3}$ for PSR B1929+10. To minimise the analysis time,
we gated and coherently dedispersed only the portion around the pulse (about
14\% of the rotation period for PSR B0950+08, and about 8\% for PSR
B1929+10). The width of the gate was chosen on the basis of the known widths of
the average pulse profiles. The widths at 10\% of the pulse peak intensity are
$\sim$8\% \& $\sim$6\% of the periods for B0950+08 \& B1929+10 respectively. The
dispersion smearing within our band for both these pulsars is much smaller than
the width of the gate.

In order to detect the signature of discrete refractive scattering events, 
we looked for
significant features in the Autocorrelation (AC) function computed from the
coherently de-dispersed voltage series. Since the delay we expect is of the
order of microseconds (or less), there was no need to compute the AC function
over a very long delay range. We computed the AC function of short stretches,
each of length 102.4 $\mu$sec ($2^{11}$ sampled points), by computing the power
spectrum and inverse fourier transforming it back, after correcting for
instrumental effects. There is also another important reason for restricting the
length of the AC functions to such a short length. PSR B0950+08 is known to show
significant short-time scale features (microstructure) in its pulse window. The
typical time scale for these features is about 175 $\mu$sec (Hankins
1972). Since it is important to minimise the contribution of this microstructure
to our AC function, we chose our transform lengths to a value of 102.4 $\mu$sec,
which is smaller than the microstructure time scale.

Since the input voltage time series is a ``real-function'', the power spectrum
computed over 20 MHz would be (real \&) symmetric about the `zero' frequency.
Any discrete feature in the AC function manifests itself as a ``fringe pattern''
(or a periodic modulation)
in this power spectrum. Since the phase of such a pattern is also of some
interest, while taking the inverse fourier transform we use only a one-sided
power spectrum (say, frequencies $\ge$0), such that the AC function thus
obtained contains both, amplitude \& phase, information (i.e. this AC function
would be a complex function) associated with any spectral modulation.

It should be emphasised here that we compute the autocorrelation of
the voltage series, and not the ``detected'' (i.e., intensity)
series. It is easy to see that the fourier transform of the latter
would correspond to a {\it fluctuation spectrum} of the intensity,
rather than the {\it radio frequency spectrum} we wish to study. This
allows us not only to reduce considerably the effect of any intrinsic
fluctuations like microstructure, but also to measure more directly
the time delays of interest.

Within the pulse window, autocorrelation functions were computed by taking
half-overlapping time sections of $2^{11}$ points. In order to improve the
signal-to-noise ratio, all these autocorrelation functions were added together
with weights given by the square of their individual signal-to-noise ratios
(which is estimated from the mean power in the power spectrum). Furthermore, we
repeat this procedure over many pulses and average all the individual
autocorrelation functions, to compute the final autocorrelation function.

\subsection{Instrumental and other spurious effects}
While computing the mean of the power spectrum, we made sure to identify all the
spurious interference signals, and to not include them in the calculation. We
replaced the spectral channel contributions affected by the narrow-band
interference by the value corresponding to the mean contribution from the
unaffected channels, so that they do not corrupt the autocorrelation function
estimation. It was very important to remove such contributions. Otherwise they
bias the mean estimations, and can also introduce strong intermodulation
products.

\subsubsection{Instrumental response function}
The estimations of two major instrumental effects which had to be compensated
for are an essential part of our analysis. The two effects are related to, (1)
the intrinsic band shape of the instrument, and (2) the {\it front-end
response}. The reason for considering the front-end (signal path from the
telescope up to the first amplifier) as a separate entity is that a very
significant fraction of the system noise, contributed by the first amplifier,
does not sample the spectral response of the front-end. This means that the
spectral response estimated on the basis of the off-pulse region of the data set
can not be used to adequately correct for the instrumental response to the
sky/pulsar signal.

We model the instrumental response function as follows:
\begin{equation}
B_{\rm on}(\nu)\;=\;\langle B_{\rm off}\rangle (\nu)\;\left[1\;+\;\left(\frac{
A_{\rm on} - \langle A_{\rm off}\rangle }{ \langle A_{\rm off}\rangle}
\right)\;D(\nu)\right]
\end{equation}
\noindent
Here, $B_{\rm on}(\nu)$ is the on-pulse bandshape as a function of frequency
(obtained for a given stretch of $2^{11}$ points), $\langle B_{\rm off}\rangle
(\nu)$ is the average off-pulse bandshape as a function of frequency, $A_{\rm
on}$ is the mean of the instantaneous on-pulse bandshape, and $\langle A_{\rm
off}\rangle$ is the mean of the average off-pulse bandshape. $D(\nu)$ is the
front-end response function.

In order to correct for the above, we pre-computed a high-signal-to-noise ratio
average off-pulse band shape function, $\langle B_{\rm off}\rangle (\nu)$. We
also pre-computed the function $D(\nu)$ by inverting the above equation and by
estimating its suitably weighted average as

\begin{equation}
D(\nu)\;=\;\frac{ \sum_{i=1}^{M} \left[\frac{B_{\rm on}(\nu)}{\langle B_{\rm
off}\rangle}-1\right] \; w_i} {\sum_{i=1}^{M} {w_i}^2}
\end{equation}
\noindent
here, $M=(N_{\rm str}\times N_{\rm pul})$, with $N_{\rm str}$ and $N_{\rm pul}$
being the number of stretches in a pulse \& the number of pulses considered for
the analysis, respectively and $w_i = [A_{\rm on} / \langle A_{\rm off}\rangle -1]$ is
a quantity proportional to the signal-to-noise ratio of the pulse in the
corresponding stretch.

\begin{figure}
\epsfig{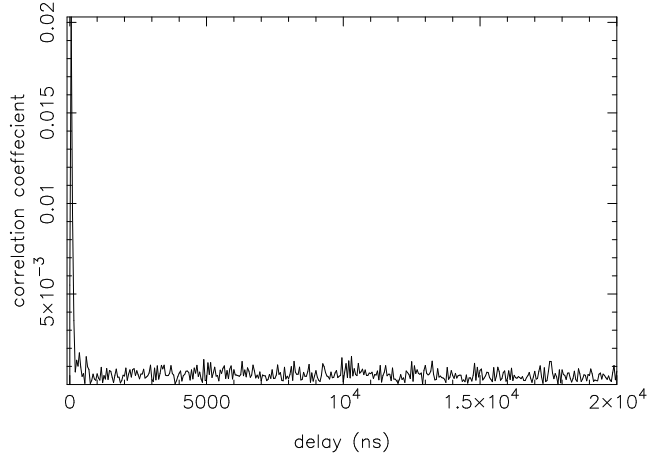}
\caption[]{Autocorrelation function computed for PSR B0950+08. This function is
based on the analysis of 632 pulses observed at a centre
frequency of 382 MHz, with 10 MHz bandwidth. See text for details.}
\label{fig:0950_ac}
\end{figure}

Then, before computing the autocorrelation functions, we compensated for the
instrumental effects to obtain 

\begin{equation}
B_{\rm on}(\nu)\;=\; \left[{B'_{\rm on}(\nu) \;-\; \langle B_{\rm off}\rangle
(\nu)} \right]/ \left[{D(\nu)\; \langle B_{\rm off}\rangle (\nu)} \right]
\end{equation}
\noindent
where, $B'_{\rm on}(\nu)$ is the uncorrected power spectrum. The corrected power
spectrum, $B_{\rm on}(\nu)$, is inverse fourier transformed to estimate the
autocorrelation as function of delay $\tau$.
The front-end response function, $D(\nu)$, was determined using the pulsar
observation itself, and a smooth functional form (with a low-order polynomial)
fitted to the estimate was used for the correction described. This ensured that any fine-scale
spectral modulation in the pulsar signal was not removed as part of the
instrumental response.

\section{Results}
We analysed 632 pulses for PSR B0950+08, and 700 pulses for PSR B1929+10. The
final AC functions are normalised according to the following procedure.

As described in section \ref{sec-analysis}, in each stretch, we have $N =
2^{11}$ samples (102.4 $\mu$sec). With the root-mean-square value (r.m.s.) of
the initial voltage time series $\sigma_{\circ}$, the r.m.s. of the noise in the
real and the imaginary parts of the normalised AC function is
$\sigma \;=\; \sqrt{{2}/{N}}$.

In our procedure, many such AC functions are added together. The number of AC
functions in either the X or the Y polarisation channel is given M as defined above.
The final AC function $\langle C(\tau)\rangle$ is computed by a weighted average given by,

\begin{equation}
\langle C(\tau)\rangle\;=\; \frac{\sum_{i=1}^{M}\;
W_{i}\;C_{i}(\tau)}{\sum_{i=1}^{M}\; W_{i}}
\end{equation}
\noindent
where $C_i(\tau)$ is the AC function for the $i^{\rm th}$ stretch and the weightage $W_i$ 
(=${w_i}^2$) is proportational to the square of the 
signal-to-noise ratio of the pulsar contribution in that
102.4$\mu$sec stretch.

The r.m.s. of the real and imaginary parts of the normalised final AC
function is $\sigma_f\;=\; {\sigma}/{\sqrt{M_{\rm eq}}}\;=\;
\sqrt{{2}/({N\times M_{\rm eq}})}$,
where $M_{\rm eq}$ is the equivalent number of ACFs averaged ($\le M$). 
If the weights for all the stretches are equal, then $M_{\rm eq} = M$.

\begin{figure}
\epsfig{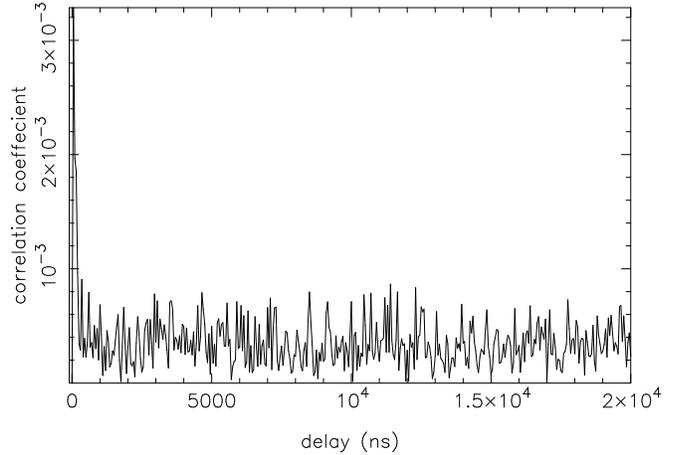}
\caption[]{Autocorrelation function computed for PSR B1929+10. This function is
based on the analysis of 700 pulses observed at a centre
frequency of 382 MHz, with 10 MHz bandwidth. See text for details.}
\label{fig:1929_ac}
\end{figure}

Figures \ref{fig:0950_ac} \& \ref{fig:1929_ac} give the normalised amplitude of
the final (average complex) AC function for PSRs B0950+08 and B1929+10,
respectively. Note that the AC function resulting from this analysis has a null
at `zero-delay', i.e.  the feature corresponding to the mean spectral
contribution is removed. As our radio frequency bandwidth is 10 MHz, in
principle, we should be able to measure any feature whose time delay is greater
than $\sim$100 ns. However, due to a finite residual in the instrumental
response which remains uncompensated despite our detailed modelling, the minimum
delay for our ACF estimation is $\sim$200 ns. As we can see, there is no
significant feature in the AC functions beyond this delay. This indicates an
absense of a) prominant discrete refractive scattering events as well as b)
non-multiple diffractive scattering along both lines of sight.

\section{Discussion}

In our observations, we have looked for fine periodic spectral structure across
a bandwidth not very much wider than the decorrelation bandwidth.  Also, the
data span used corresponds to an interval smaller than the decorrelation time
for diffractive scintillations of both pulsars. Thus our analysis uses data over
at most a couple of diffractive scintles, and is therefore less affected by any
possible decoherence due to variation of the diffractive phase from scintle to
scintle.  Considering the possibility that the spectral modulation (which we
have searched for) could drift across the spectrum with time (although the drift
is expected to be small over our data span), we have also examined the average
of ACF amplitudes alone (instead of its vector equivalent) so as to avoid
possible `dephasing' of any otherwise detectable ACF feature.  Again, no
significant feature was apparent.

We interprete our {\it null} result as indicating an absense of strong
refraction gradients across the scattering screens. Our results also suggest
that the diffractive scattering is rather `multiple', i.e. the diffraction image
consists of a large number of `speckles' or sub-images.  While the former
indication can be understood simply as the result of a non-steep spectrum of
density irregularities with an `inner scale' much smaller than the Fresnel
scale, we attempt to quantify our conclusion using the following simple-minded
picture.

Crossing of refracted ray bundles is a necessary condition for multiple imaging
(see Cordes et al 1986 for more details). This condition is not met if the
differential refraction across a given transverse scale is less than its angular
size as seen from the observer's location.  Let us consider a refracting screen
mid-way along a pulsar sight-line, and assume that two wavefronts, one of which
is refracted significantly more (by, say, an angle $\theta$) than the other,
describe the refractive effect of the scattering screen over a transverse
separation `$a$'.  For a difference $\Delta DM$ in the electron column density
sampled by the two paths, the two wavefronts will cross if $\Delta DM$ $\ge$
$(8\pi/Lr_e) (a/\lambda)^2$, where $r_e$ is the classical electron radius, $L$
is the distance to the pulsar, and $\lambda$ is the wavelength of observation.
For the situation described here, it is easy to show that the geometric delay
difference will be equal to the differencial dispersion delay corresponding to
$\Delta DM$.

Figure 3 shows the above mentioned limits on column density contrast 
$\Delta DM$ (in units of pc cm$^{-3}$) associated with large-scale irregularity
as a function of transverse scale `$a$' (in AU) for the two pulsars,
B0950+08 (solid line) and B1929+10 (`dash' line). The distances to these two
pulsars are assumed to be 120 pc and 170 pc, respectively. The vertical axis
(right side) indicates the associated relative delay (including the
dispersion contribution) for comparison with our measurements.  The horizontal
axis (top side) is also marked with relative epochs (for the line of sight to
traverse a mid-way transverse distance of $a$) corresponding to a proper motion 
of 100 km/s. It would be instructive to compare these limits on the structure function
of dispersion measure when suitable direct measurements of changes in DM in the
two directions may become available in future. It is also worth mentioning that
the characteristic delay associated with diffractive scattering of the two cases
(B0950+08 \& B1929+10) are 4.7 \& 0.08 $\mu$sec, respectively, and would
translate to scattering disk size of 0.8 \& 0.12 AU, respectively. The Fresnel
scale, for comparison, is $\sim 10^{-2}$ AU.

\begin{figure}
\epsfig{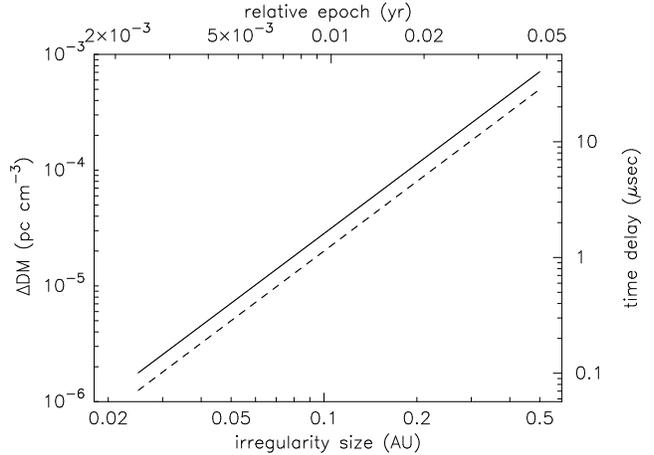}
\caption[]{Upper bounds to the structure function of dispersion measure. The
solid line is for PSR B0950+08, and the `dash' line is for B1929+10. See text
for details.}
\label{fig:ne_vs_a}
\end{figure}

It is worth mentioning that the power spectrum of irregularities in the
direction of B1929+10 appears relatively steep (Bhat, Gupta \& Rao 1999) and
hence the refractive effects can be expected to be dominant, relatively
speaking.  Also, the multipath scale (defined by the apparent size of the
scatter-broadened image) when viewed in units of the Fresnel scale is an order
of magnitude smaller for B1929+10 than that for B0950+08, indicating
correspondingly weaker and less multiple scattering.  However, it is not clear
as to how much of this apparent difference is due to the possible differences in
the fractional distance to the scatterer (i.e. the scatterer location may be at
one end in the case of B1929+10 instead of mid-way as we have assumed).
It should be emphasised here that the refractive effects are highly episodic. So
our ({\it null}) results on these two pulsars, and the implied constraints (as in
Figure \ref{fig:ne_vs_a}), while being indicative, should be viewed as applying 
strictly only to the epochs of our observations.  

To summarise, radio signals emitted by very near-by sources may be scattered by
discrete refractive irregularities that alter the diffraction patterns
significantly, or may undergo `non-multiple' diffractive scattering.  Finding
distinct spectral signatures of such events allows us to study the size and the
density contrast of such large-scale irregularities. We have looked for such
signatures in the signals from PSR B0950+08 and PSR B1929+10, but found no
corresponding ACF feature significantly above the noise threshold ($7\sigma$) of
$2\times 10^{-3}$ and $1.3\times 10^{-3}$, respectively. The absence of any such
signature/event in the observed data suggests useful limits on the structure
function of dispersion measure corresponding to a wide range of refractive
scales in the two lines of sight (as sampled at the epoch of our observations).   

\section*{Acknowledgements}
The software code for coherent de-dispersion was developed by Dipankar
Bhattacharya, along with one of the authors (RR). We thank our referee,
Jim Cordes, for valuable critical comments that have helped us in 
improving the paper significantly.


\begin{thebibliography}{}
\bibitem{} Bhat N.D.R., Gupta Y., Rao A. P. 1999, ApJ, 514, 249
\bibitem{} Blandford R. D., Narayan R., 1985, MNRAS, 213, 591
\bibitem{} Blandford R. D., Narayan R., Romani R. W., 1984, A\&A, 5, 369
\bibitem{} Cordes J. M., 1986, ApJ, 311, 183
\bibitem{} Cordes J. M., Pidwerbetsky A., Lovelace R. V. E. 1986, ApJ, 310, 737
\bibitem{} Cordes J. M. \& Wolszczan A. 1986, ApJ, 307, 27
\bibitem{} Goodman J. \& Narayan R. 1989a, MNRAS, 231, 97
\bibitem{} Goodman J. \& Narayan R. 1989b, MNRAS, 238, 995
\bibitem{} Hankins T. H., 1971, ApJ, 169, 487
\bibitem{} Hankins T. H., 1972, ApJ, 177, 11
\bibitem{} Lee L. C., Jokipii J. R., 1975, ApJ, 201, 532
\bibitem{} Lee L. C., Jokipii J. R., 1976, ApJ, 206, 735
\bibitem{} Scheuer P. A. G., 1968, Nature, 218, 920
\end{thebibliography}
\end{document}